\documentclass[journal]{IEEEtran}
\ifCLASSINFOpdf
\else
\fi
\usepackage[dvipsnames]{xcolor}
\usepackage{tikz}

\usepackage{xspace}
\newcommand{\eg}{\textit{e.g.},\xspace}
\newcommand{\ie}{\textit{i.e.},\xspace}
 
\usepackage[cmex10]{amsmath}
\usepackage{booktabs}     

\usepackage{stfloats}
\usepackage{graphicx}
\usepackage{color}
\usepackage{placeins}
\usepackage{float}
\usepackage{tabularx,colortbl,multirow}

\usepackage{multirow}
\usepackage{graphicx}
\usepackage{subcaption}
\usepackage{enumerate}
\usepackage{enumitem}
\usepackage{amsmath,amssymb,amsfonts,amsthm,tikz,caption}

\usepackage{mathtools}

\usepackage{bm}

\renewcommand\eqref[1]{(\ref{#1})}
\graphicspath{{images/}}
\usepackage{tikz}
	\usetikzlibrary{shapes,decorations,decorations.pathreplacing}
	\usetikzlibrary{arrows,shapes,positioning,calc,fadings,patterns}
	\usetikzlibrary{decorations,decorations.pathreplacing,decorations.pathmorphing}
\usepackage{tikz-3dplot}

\usepackage{booktabs}     
\usepackage{threeparttable}
\usepackage{amssymb}
\usepackage[style=ieee] {biblatex}
\bibliography{GEM_TAP} 

\makeatletter
\newcommand{\vast}{\bBigg@{2.5}}
\makeatother

\usepackage{algorithm}%
\usepackage{algorithmicx}%
\usepackage{algpseudocode}%
\newtheorem{theorem}{Theorem}%
\newtheorem{proposition}[theorem]{Proposition}

\usepackage{siunitx}

\begin{document}
%
\title{Solving Maxwell’s Equations with Non-Trainable
Graph Neural Network Message Passing
}

%

\author{Stefanos~Bakirtzis, Marco Fiore,~Jie~Zhang, and Ian Wassell 
\thanks{The work of Stefanos Bakirtzis is supported by the Onassis Foundation and the Foundation for  Education and   European Culture.}
\thanks{Stefanos Bakirtzis,  and Dr. Ian Wassell are with the Department of Computer Science and Technology, University of Cambridge,  Cambridge, CB3 0FD, United Kingdom (e-mail: ssb45@cam.ac.uk;   ijw24@cam.ac.uk).}

\thanks{Dr. Marco Fiore is with the IMDEA Networks Institute, Madrid, Spain    (e-mail: marco.fiore@imdea.org).}

\thanks{Professor Jie Zhang is with the Department of Electronic and Electrical Engineering, University of Sheffield,  Sheffield, S10 2TN, United Kingdom (e-mail: jie.zhang@sheffield.ac.uk).}


}

\maketitle


\begin{abstract}


 Computational electromagnetics (CEM) is employed to numerically solve Maxwell’s equations, and it has very important and practical applications across a broad range of disciplines, including biomedical engineering, nanophotonics, wireless communications, and electrodynamics. The main limitation of existing CEM methods is that they are computationally demanding. Our work introduces a leap forward in scientific computing and CEM by proposing an original solution of Maxwell’s equations that is grounded on graph neural networks (GNNs) and enables the high-performance numerical resolution of these fundamental mathematical expressions. Specifically, we demonstrate that the update equations derived by discretizing Maxwell’s partial differential equations can be innately expressed as a two-layer GNN with static and pre-determined edge weights. Given this intuition, a straightforward way to numerically solve Maxwell’s equations entails simple message passing between such a GNN’s nodes, yielding a significant computational time gain, while preserving the same accuracy as conventional transient CEM methods. Ultimately, our work supports the efficient and precise emulation of electromagnetic wave propagation with GNNs, and more importantly, we anticipate that applying a similar treatment to systems of partial differential equations arising in other scientific disciplines, e.g., computational fluid dynamics, can benefit computational sciences.


\end{abstract}


%
\IEEEpeerreviewmaketitle

\vspace{ - 2mm }

\section{Introduction}
%
%
%
%

The solution of partial differential equations (PDEs)  is of paramount importance in scientific computing and computational physics, as these mathematical expressions describe a plethora of physical phenomena \cite{Book_PDE}.  
Maxwell's equations are among the most important and well-known sets of PDEs and they represent the governing laws of electromagnetism, describing the propagation of electromagnetic waves and their interaction with the physical environment. Despite the fidelity of Maxwell's equations, exploiting them to retrieve an analytical solution for the electromagnetic field components is infeasible for most real-world problems. Instead, computational electromagnetics (CEM) provides numerical solutions to Maxwell's equations and it has been applied in various fields including biomedical engineering, geophysics, nanophotonics, wireless communications, and antenna and radio frequency component design  \cite{CEM, FDTD_Bio}. Existing CEM methods can be categorized based on whether they solve Maxwell's equations in a differential or integral form, or according to whether the solution is retrieved in the frequency or the time domain \cite{App2_Photonics}.  Some of the commonly used CEM full-wave analysis models include the finite element method \cite{hughes2012finite}, the method of moments  \cite{gibson2021mom}, and the finite-difference time-domain (FDTD) method \cite{Taflove_FDTD}. All these approaches necessitate high computational resources, which emerge as their main bottleneck.  Among them, FDTD constitutes the most rigorous technique for high-fidelity transient (time-domain) solutions to Maxwell's equations, while it also enables broadband frequency domain responses to be acquired via a Fourier transformation of the sampled waveforms \cite{FDTD_Nature_Methods}.

The computational burden in the FDTD method stems from the discretization of Maxwell’s equations both in space and time via a centered finite-difference approximation \cite{Gedney_FDTD}. Once discretized,  the electromagnetic fields at a given point in space and time are computed iteratively through a set of numerical equations. The values of the spatial and time discretization steps used in the centered finite-difference approximation are subject to constraints that must be met in order to satisfy certain stability and numerical dispersion requirements \cite{Gedney_FDTD, Taflove_FDTD}. This requirement yields spatial step sizes that are a subdivision of the wavelength, and values of the time step that are smaller than the spatial step by a factor proportional to the speed of light. Inevitably, simulating electromagnetic propagation in large physical environments and very high frequencies can lead to overextended simulation times and exhaustive memory requirements, hindering the wide applicability of such models.

In this work, we propose an expedient CEM framework for transient solutions to Maxwell's equations based on graph neural networks (GNNs), referred to as GEM. GEM is founded on a very simple, yet fundamentally innovative and elegant observation: \textit{the spatial locality of GNNs, \ie the fact that graph node values depend on their own previous values and the values of their neighboring graph nodes, is semantically aligned with the mechanisms underpinning the FDTD method, where the electromagnetic field values are iteratively computed based on the previous time step and the neighboring electromagnetic field components}. From this observation springs an equivalency between the discretized Maxwell's equations and a specific GNN topology. Hence, a numerical solution to discretized Maxwell's equations can be achieved via a fixed-weight, non-trainable GNN, which we call GEM, that can faithfully replicate the results of the full-wave numerical analysis method.

GEM consists only of two message-passing neural network (MPNN)  layers \cite{MPNN} with static and pre-determined edge weights, one responsible for updating the electric fields and the other updating the magnetic fields in the same manner as in the FDTD method. However,   exploiting the native graphic processing unit (GPU) implementations of GNNs \cite{Py_Geometric}, GEM can solve Maxwell’s equations up to 40 times faster than conventional FDTD based on central processing unit (CPU) parallelization while yielding exactly the same results; 
interestingly, we also observe that GEM is \textit{at least} twice as fast as state-of-the-art FDTD implementations that exploit advanced optimizations and parallelization, \eg grid chunking \cite{Taflove_FDTD}, even if our solution does not yet adopt those same techniques. Contrary to some recent trends in the literature ~\cite{ML_FDTD_1, ML_FDTD_2, ML_FDTD_3, ML_FDTD_4, ML_FDTD_5, GNNs_FDTD}, it is shown that it's not necessary to identify appropriate input features or deep learning architectures aimed at training data-driven CEM models. Instead, FDTD equations can be intrinsically transformed into a GEM model that fundamentally outperforms the legacy implementation of the full-wave solver in terms of computational performance, without any need for training.

Furthermore, GEM  can be exploited in the context of physics-informed neural networks for the solution of PDEs \cite{Dif_Eq_Karniadakis, karniadakis2021physics}. Indeed, there is a plethora of research on data-driven computational sciences applied in different domains such as chemistry, biology, fluid dynamics, and neuroscience \cite{Deep_Learning_Chemistry, Deep_Learning_Biology, Deep_Learning_Fluid, Deep_Learning_NeuroScience}. Instead of pursuing solely data-driven approaches where neural networks need to be trained to solve a system of PDEs arising in physics, we foster the adoption of graph-driven approaches, where a GNN with properly defined edge weights can rigorously solve the exact target PDE system. Such models can effectively replace conventional scientific computing techniques since they provide identical results and are substantially faster. Consequently, non-linearities and learnable components can be included in the graph-driven solvers, to either account for real-world measured data (in cases where the fidelity of partial differential equation is not guaranteed) or to operate outside the stability limits that pose strict requirements for the spatial and temporal steps of conventional scientific computing techniques. 

\section{Related Work}

To accelerate the numerical solution of the discretized Maxwell equations, domain decomposition parallelization schemes have been studied, aiming at partitioning the computational domain into several subdomains \cite{FDTD_GPU_MPI, FDTD_MPI_Parition}. The solution of the update equations at each subdomain is allocated to a different processing unit and subsequently, the individual solutions are combined to derive a solution for the entire domain. Typically, the decomposition is programmed so that the individual solutions are implemented on different processing units, thus speeding up the iterative update process \cite{Accelerate_GPU, Taflove_FDTD, CUDA_FDTD, FDTD_GPU_Acc}. All these approaches entail cumbersome transformations of the legacy discretized Maxwell's equations that eventually curbed their adoption  \cite{ML_FDTD_5}.

Lately, and in light of the recent advancements in artificial intelligence and especially in deep learning (DL), the idea of data-driven CEM has become increasingly popular. Such models proffer a unique opportunity to overcome the computational efficiency limitations associated with full-wave analysis \cite{ML_CEM, ML_CEM_2}. Specifically, low computational complexity DL algorithms can be trained offline to learn the
underlying laws of electromagnetism, approximated through numerical methods, and they can be subsequently employed to replicate the results of these methods for new scenarios.  For instance, approximate electromagnetic methods, such as ray tracing, have been coupled with DL algorithms to forge generalizable data-driven propagation models \cite{Aris_CNN, EM_DeepRay, DeepRay,  CNN_2020, levie2021radiounet}. Similarly, time-invariant full-wave methods, such as finite elements and method of moments, have also been coupled with DL to emulate steady-state results \cite{MoM_ML, FEM_Deep, FEM_Deep_2}. However, all these models are static in time, whereas transient numerical solutions to Maxwell’s equations entail simulating wave propagation simultaneously in space and time. Thus, DL algorithms entwined with a time-domain method should be able to capture both the spatial dependency of the electromagnetic field intensity and its variation over time.

In this direction, the use of convolutional neural networks (CNNs) and sequence models, such as long short-term memory networks has been explored \cite{ML_FDTD_1, ML_FDTD_2, ML_FDTD_3, ML_FDTD_4, ML_FDTD_5}. Despite being able to capture spatiotemporal correlations, the scalability of such models as standalone solvers of Maxwell's equations is innately limited. Indeed, all these DL architectures invariably exhibit low versatility since conventional CNNs are can be applied to fixed-size input spatial domains.  More recently,  GNNs were employed for the inference of electromagnetic fields \cite{GNNs_FDTD, GNN_Bakir}, due to their potential to respect the topological properties of the geometry, and exploit the intrinsic graph-like structures that appear in CEM problems.  Unlike CNNs which are designed to operate with structured data, GNNs can handle varying size topologies with arbitrary structure. In \cite{GNNs_FDTD, GNN_Bakir}, the  GNN   received as input the electromagnetic properties of the simulated domain and the field components of the previous time steps and then inferred the electromagnetic field evolution. Although the trained data-driven model could predict the fields at the next time iteration with a small error, using it as a standalone CEM model was infeasible; indeed, the authors observed a fast error accumulation leading to a severe deterioration of the prediction accuracy and a breakdown of the solver after just a few iterations.

\section{Solving to Maxwell's Equations with Finite Differences} \label{Sec:Disc}

\subsection{Maxwell's Equations}

Maxwell's equations constitute the foundations of classical electromagnetism  \cite{jackson1999classical} and they portray effectively all electromagnetic phenomena.  The third and fourth  equations, also known as the Ampere and Faraday laws, capture the impact of a time-varying electric field  on the  magnetic field and vice versa:

  \begin{equation}
     \nabla \times \mathbf{H} - \epsilon \frac{\partial\mathbf{E}}{\partial t} = \mathbf{J}
      \label{eq:Maxw_H}
 \end{equation}
 
  \begin{equation}
     \nabla \times \mathbf{E} + \mu \frac{\partial\mathbf{H}}{\partial t} = - \mathbf{M}
      \label{eq:Maxw_E}
 \end{equation}

 \noindent where $\mathbf{E}$ and $\mathbf{H}$ represent the electric and magnetic field intensity vectors, respectively, whilst $\mathbf{J}= \sigma \mathbf{E}$ and $\mathbf{M} = \sigma^* \mathbf{H}$   denote the electric and magnetic current density vectors,  and $\epsilon$,  $\mu$, $\sigma$, and $ \sigma^*$  are the permittivity, the permeability, the electric and the magnetic conductivity of the medium, respectively. Assuming a non-magnetic material the partial differential equations emerging from~(\ref{eq:Maxw_H}) and (\ref{eq:Maxw_E}) for the case of a  2D transverse electric (TE)  polarization are:

\begin{equation}
       \hspace{3mm }
    \sigma E_y+ \epsilon \frac{\partial E_y}{\partial t} = \frac{\partial H_x}{\partial z}  - \frac{\partial H_z}{\partial x}  
      \label{eq:TE_1}
 \end{equation}

\begin{equation}
  \mu \frac{\partial H_x}{\partial  t} = \frac{\partial E_y}{\partial z}  
      \label{eq:TE_2}
 \end{equation}

\begin{equation}
   \hspace{3.45 mm }
  \mu \frac{\partial H_z}{\partial  t} = - \frac{\partial E_y}{\partial x}  
      \label{eq:TE_3}
 \end{equation}

\subsection{The Finite-Difference Time-Domain method}

Maxwell's equations are intractable for most real-world problems, which has motivated the development of  CEM methods. A landmark and breakthrough in transient CEM was Yee's algorithm  \cite{Yee_Paper} which suggested partitioning the continuous space into a grid of discretized spatial cells and then ordering the electromagnetic field components on the sides of each cell in a way that satisfies Ampere's and Faraday's law. Consequently,  numerical update equations of the electromagnetic fields are derived by employing a centered finite-difference scheme to approximate the spatial and temporal derivatives  of a field,  ${f}$, around a point $x_0$  as follows:

\begin{equation}
   \hspace{3.45 mm }
   \partial f(x)/ \partial x \bigl\rvert_{x = x_0}  \approx \frac{f\left(x + \displaystyle \frac{\Delta x}{2}\right) - {f}\left(x - \displaystyle \frac{\Delta x}{2}\right) }{ \Delta x}
      \label{eq:Centered_Difference}
 \end{equation}

 \noindent Applying this treatment to the equations of the TE polarization, (\ref{eq:TE_1}) - (\ref{eq:TE_3}), centered around  the points   $[x, z, t] = [i+\frac{1}{2}, j+\frac{1}{2}, n + \frac{1}{2}]$, yields the following update equations:

\begin{multline}
    E_{y_{i+\frac{1}{2}, k+\frac{1}{2}}}^{ n+1} = \frac{A_-}{A_+}  E_{y_{i+\frac{1}{2}, k+\frac{1}{2}}}^{ n} +  \frac{
  \Delta t} {A_+ \cdot \epsilon_{y_{i+\frac{1}{2}}, k+\frac{1}{2}}} \cdot  \\\ \left( \frac{H_{x_{
  i+\frac{1}{2}, k+1}} ^ { n+ \frac{1}{2} } - H_{x_{i+\frac{1}{2}, k}}^ {n + \frac{1}{2}}}{\Delta z}  -      \frac{H_{z_{i+1, k+\frac{1}{2}}}^ { n+ \frac{1}{2} } - H_{z_{i, k+\frac{1}{2}}}^ {n +\frac{1}{2}}}{\Delta x} \right) 
 \label{eq:Update_Ey}
\end{multline}

\begin{equation}
    H_{x_{i+\frac{1}{2}, k}}^{ n+\frac{1}{2}} =   H_{x_{i+\frac{1}{2}, k}}^{ n-\frac{1}{2}} + \frac{ \Delta t}{\Delta z \cdot \mu_{i+\frac{1}{2}, k} } \left( E_{y_{
  i+\frac{1}{2}, k+\frac{1}{2}}} ^ { n } - E_{y_{i+\frac{1}{2}, k-\frac{1}{2}}}^ {n }  \right) 
      \label{eq:Update_Hx}
 \end{equation}

\begin{equation}
    H_{z_{i, k+\frac{1}{2}}}^{ n+\frac{1}{2}} =   H_{z_{i, k+\frac{1}{2}}}^{ n-\frac{1}{2}}- \frac{ \Delta t}{\Delta x  \cdot \mu_{i, k+\frac{1}{2}}} \left( E_{y_{
  i+\frac{1}{2}, k+\frac{1}{2}}} ^ { n } - E_{y_{i-\frac{1}{2}, k+\frac{1}{2}}}^ {n }  \right) 
      \label{eq:Update_Hz}
 \end{equation}

 \noindent where $i, j$ and $n$ are non-negative integers, $A_{-,+} = 1 \mp \displaystyle \frac{\sigma_{i + \frac{1}{2}, k+\frac{1}{2}} \Delta t}{2 \displaystyle \epsilon_{i+\frac{1}{2}, k +\frac{1}{2}}}$,  $\Delta t$ is the time step, and $\Delta x$ and $\Delta z$ are the spatial steps in the horizontal and vertical direction, respectively. Similar equations can be straightforwardly derived for the transverse magnetic (TM) polarization or the 3D case. As can be seen,  the electromagnetic field components are sampled at spatial locations offset by half a pixel, and they are arranged into a mesh as shown in Figure~\ref{fig:GEM_Framework}b. Also,  let $n$ be an integer,  by convention, the electric field components are evaluated at time steps that are multiples of $n$ and the magnetic field components at time steps that are multiples of $(n+\frac{1}{2})$.


The update scheme entails first evaluating each magnetic field component using the neighboring electric field node values at the current time step, and the values of the respective magnetic field at the previous time step. Then,  the computed $H_x$ and $H_z$ values along with the values of $E_y$ at the previous time step are employed to estimate the electric fields at the next iteration, which are consequently used to update again the magnetic fields, and so on. Ultimately, the evolution of the electromagnetic fields over space and time is posed as a two-step iterative procedure. 
 
\subsection{Absorbing Boundary Conditions}

The solution of Maxwell's equations via (\ref{eq:Update_Ey}) -- (\ref{eq:Update_Hz}) will introduce numerical artifacts, and in particular artificial reflections, when the electromagnetic waves reach the boundary of the grid. To overcome this limitation and terminate the grid successfully, various types of absorbing boundary conditions have been proposed in the literature \cite{Taflove_FDTD}. Among them, the perfectly matched layer  (PML)  constitutes the most rigorous and expedient grid termination method \cite{PML_2D}. According to this approach, the main grid is extended by placing at the end of the mesh an additional lossy layer of cells. Thus, the propagating waves in the PML region gradually attenuate and ultimately completely vanish. The parameters of the PML area are selected such as that the impedance of the additional lossy cells matches perfectly the impedance of the main grid at all the angles and frequencies of an incident wave, thus ensuring that there are no spurious reflections due to the material parameter discontinuity. 

For the implementation of the PML in the TE case, the electric field component is split into two subcomponents denoted by $E_{yx}$ and $E_{yz}$, where $E_y =  E_{yx} + E_{yz}$. Furthermore, at the added lossy cells the material is assumed to be magnetic and the relationship $\sigma/\epsilon =  \sigma^*/\mu$ should hold (note that the magnetic conductivity affects only the PML cells). Then,  taking into consideration the application of the split field formulation,   Maxwell's differential equations  (\ref{eq:TE_1}) -- (\ref{eq:TE_3})  can be replaced by four equations \cite{PML_2D}: 

\begin{equation}
       \hspace{3mm }
    \sigma_x E_{yx}+ \epsilon \frac{\partial E_{yx}}{\partial t} =    - \frac{\partial H_z}{\partial x}  
      \label{eq:TE_1_PML}
 \end{equation}

\begin{equation}
       \hspace{3mm }
    \sigma_z E_{yz}+ \epsilon \frac{\partial E_{yz} }{\partial t} = \frac{\partial H_x}{\partial z}     
      \label{eq:TE_1_1_PML}
 \end{equation}

\begin{equation}
\sigma^*_x H_{x}+  \mu \frac{\partial H_x}{\partial  t} = \frac{\partial(E_{yx} + E_{yz})}{\partial z}  
      \label{eq:TE_2_PML}
 \end{equation}

\begin{equation}
   \hspace{3.45 mm }
 \sigma^*_z E_{z}+ \mu \frac{\partial H_z}{\partial  t} = - \frac{\partial(E_{yx} + E_{yz})}{\partial x}  
      \label{eq:TE_3_PML}
 \end{equation}

\noindent where  $\sigma_x$, and $ \sigma^*_x$ indicate the existence of  electric and magnetic conductivity towards the $x$ direction at the boundary of the grid, whilst   $ \sigma_z$ and $\sigma^*_z$ denote  the same in the $z$ direction. The four split equations (\ref{eq:TE_1_PML}) -- (\ref{eq:TE_3_PML}) can be discretized to derive four update equations, following the same methodology described in the previous subsection. 

\section{A Graph-Driven Solution to Maxwell Equations} \label{Sec:GEM}

\subsection{Equivallent Graph Neural Network Representation}

\begin{figure*}[t]
\centering
    {\includegraphics[width = 1.7\columnwidth]{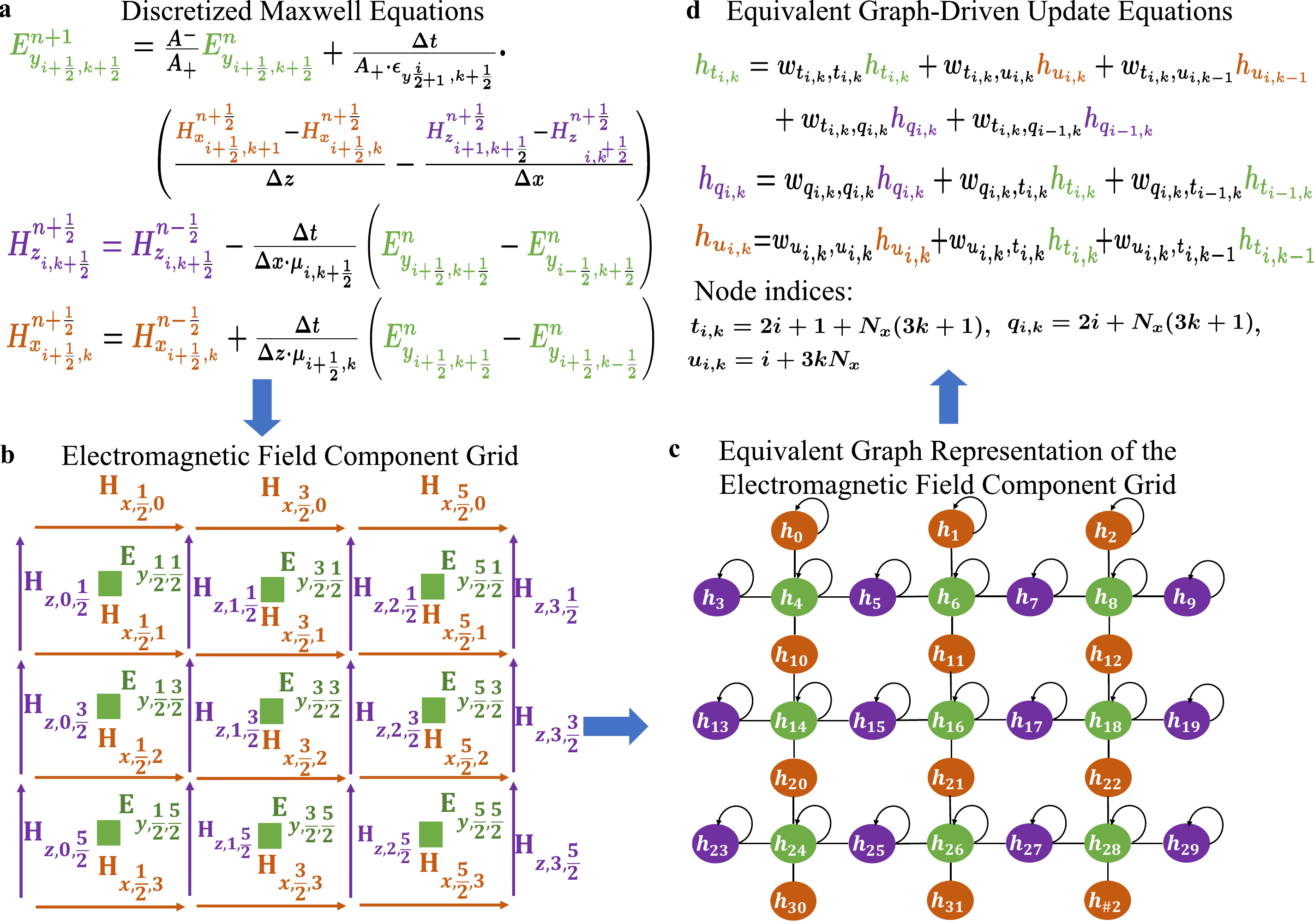}}
\caption{{Graph-driven solver of Maxwell's Equations; the $E_y$, $H_z$, and $H_x$ components are shown in green, purple, and dark orange color. \textbf{a}, Discretized TE Maxwell equations along with  \textbf{b}, a corresponding example 2D TE grid representation of the electromagnetic field component arrangement in space. \textbf{c}, The grid implicitly assumes an equivalent graph topology, and consequently,  \textbf{d}, the electromagnetic field update equations can be expressed in a graph-driven manner via mathematical operations between the graph nodes.}}
\label{fig:GEM_Framework}
\end{figure*}

The discretized Maxwell equations enshrine an underlying connection between the electromagnetic fields that becomes more evident by inspecting the layout of the electromagnetic field components in  Fig.~\ref{fig:GEM_Framework}b: \textit{the grid of magnetic and electric field nodes resembles a graph whose vertices correspond to the electromagnetic field components appearing in the discretized Maxwell equations}. Based on this intuition, by properly arranging the edge connections between the graph nodes and the corresponding edge weights, one can forge a network of neurons such that the exchange of information between the nodes faithfully reenacts the iterations leading to the solution of the discretized Maxwell equations. Specifically, the electric field nodes have loop connections to themselves as well as direct connections to their neighboring magnetic field nodes, whilst the magnetic field nodes are connected to the adjacent electric fields and they also have a self-loop connection, as shown indicatively in  Fig.~\ref{fig:GEM_Framework}c for the TE polarization case. For each connection, suitable edge weights can be straightforwardly derived by observing the counterpart coefficients of the respective field components appearing in the discretized Maxwell equations, as also illustrated in   Fig.~\ref{fig:GEM_Framework}a and Fig.~\ref{fig:GEM_Framework}d.

Then, the two-step iterative numerical solution to Maxwell's equations can be achieved through a two-layer GNN comprising two  MPNNs \cite{MPNN}. In MPNNs, the graph node values are updated through a message and a readout phase, both implemented via learned and differentiable functions. In the message-passing phase, the values of each node, $v$, are updated based on the information accrued from its graph neighborhood  via learnable non-linear operations, while the readout phase estimates a learnable feature vector for the entire graph. In GEM,  we argue that to solve differential equations the message function does not need to be learnable since the weighting coefficients can be readily deduced from the discretized equations, while the readout function can be omitted. 



 Overall, the discretization of Maxwell's equations in space and time creates a grid of nodes that intrinsically assumes a graph structure, and 
a straightforward way to numerically solve Maxwell’s equations entails simple message exchange between the graph nodes. Based on this fundamental insight, and without loss of generality (in the sense that via duality one can derive an equivalent scheme for the transverse magnetic case, or easily extend it to 3D), the following propositions are formally articulated for the TE case:
 
\begin{proposition}

The discrete electromagnetic field nodes of an $N_x \times N_z$ FDTD grid,  providing a numerical solution to Maxwell's equations, can be perceived as a directed graph $\mathcal{G}(\mathcal{V}, \mathcal{E})$, with  $3 \times N_x \times N_z$ nodes. For the set of nodes $\mathcal{V} = v_0, v_1, ... v_{3 \times N_x \times N_z-1}$, a node index equal to $t_{i,k}= 2i + 1+ N_x(3k+1)$ represents electric field components, whereas indices equal to $u_{i,k}= i + 3kN_x$ and $q_{i,k}= 2i +  N_x(3k+1)$ correspond to the $H_x$ and $H_z$ components, respectively ( $0 \leq i \leq Nx-1 $ and $0 \leq i \leq Nz-1$). An edge $e_{i,j} = (v_i, v_j) \in \mathcal{E}$ designates that there is a connection between two electromagnetic nodes $v_i$ and $v_j \in \mathcal{V}$. The set of edges emanating from an electromagnetic node $v$ is called neighborhood,  $N(v) = \{ v' \in \mathcal{V} | (v, v') \in \mathcal{E}\}$. In GEM, $\mathcal{E}$ comprises the neighborhoods of the electric and magnetic field components, defined as follows: (i) for the electric field nodes, $N(v_{t_{i,k}}) =  \{ v_{t_{i,k}},   v_{u_{i,k}},  v_{u_{i,k-1}},  v_{q_{i,k}},  v_{q_{i-1,k}} \} $, (ii)  for the $H_x$  nodes,  $N(u_{t_{i,k}})  =  \{  v_{u_{i,k}},  v_{t_{i,k}},  v_{t_{i,k-1}} \}$, and  (iii) for the $H_z$  nodes  $N(v_{q_{i,k}})  =  \{  v_{q_{i,k}}, v_{t_{i,k}}, v_{t_{i-1,k}} \}$. The edge weights are then straightforwardly derived from (\ref{eq:Update_Hx})--(\ref{eq:Update_Ey}), \eg for an edge $ e_{t_{i,k},t_{i,k}}= (v_{t_{i,k}}, v_{t_{i,k}})$ the corresponding weight is equal to  $ w_{t_{i,k},t_{i,k}}= A_-/A_+$.

\end{proposition}

\begin{figure*}[t]
\centering
    {\includegraphics[width = 1.6\columnwidth]{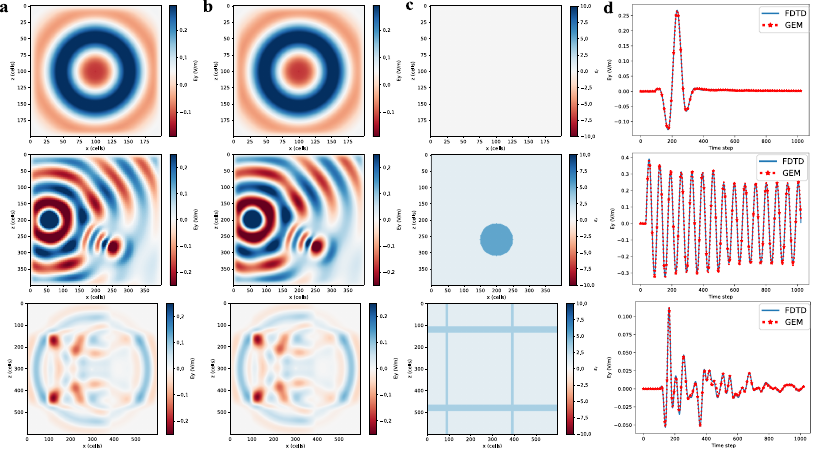}    }
\caption{Computational emulation of electromagnetic wave propagation for different use cases; \textbf{a} shows a snapshot of the electric field $E_y$  evolution simulated via GEM, whilst \textbf{b} depicts the snapshot at the same time instance as computed with FDTD. The simulated domain is shown in \textbf{c}, whereas \textbf{d} illustrates the waveform of the $E_y$ component sampled at a random point of the simulated grid. }
\label{fig:Evaluation_Results}
\end{figure*}

\vspace{-2mm}

\begin{proposition}

The nodes of $\mathcal{G}(\mathcal{V}, \mathcal{E})$ form a network of neurons, whose hidden states, $h_l$, depict the electromagnetic field evolution over space and time. Updating the discretized Maxwell  equations is equivalent to a two-step message-passing process between the network neurons. During the first step, messages flow towards $h_{u_{i,k}}$ and $h_{q_{i,k}}$ from neighboring nodes, and the new hidden state values are computed as  $h_l = \sum_{l_n \in N(v_l)} w_{l,l_n}h_{l_n}, v_l \in \{U, Q\}$, with $U, Q$ being the set of $H_x$ and $H_z$ nodes, respectively. In the second step, the neurons updated earlier remain constant and messages are conveyed towards the electric field components, $h_{t_{i,k}}$, using the same update formula, but with  $v_l \in {T} $, where $T$ is the set of $E_y$ nodes.

\end{proposition}

Propositions 1 and 2 compose GEM, which is conceived as a specialised case of the general MPNN module, where the edge weights are tuned in advance without training according to the physical problem. Conspicuously, GEM benefits from the native GPU implementation of MPNNs. Furthermore, while the message-passing and the readout functions are typically implemented via computationally expensive fully connected layers, GEM permits a reduction in the complexity of the MPNN via the use of domain knowledge that allows a proper pre-determined message-passing function to be selected and the readout function to be eliminated. Ultimately, the solution of Maxwell equations is elegantly derived by a fast matrix multiplication between edge weights and the graph node hidden states, followed by
an aggregation operation.

\vspace{-3mm}

\subsection{Graph-Driven Solution with Absorbing Boundaries}

In the presence of the absorbing boundaries, the principles of GEM remain the same with the only difference being that the graph node number is $4 \times N_x \times N_z$, and the edges are adapted to take into account the additional field components introduced due to the split field formulation. Specifically,    node indices equal to $t_{x_{i,k}}= 3i + 1+ N_x(4k+1)$  and  $t_{z_{i,k}}= 3i + 2+ N_x(4k+1)$ represents the $E_{yx}$ and  $E_{yz}$   components, while the indices  $u_{i,k}= i + 4kN_x$ and $q_{i,k}= 3i +  N_x(4k+1)$ are linked to the $H_x$ and $H_z$ components, respectively ( $0 \leq i \leq Nx-1 $ and $0 \leq i \leq Nz-1$). Additionally, the neighborhoods of the electric and magnetic field components are now changed as follows: (i) for the $E_{yx}$ component, $N(v_{t_{x_{i,k}}}) =  \{ v_{t_{x_{i,k}}},     v_{q_{i,k}},  v_{q_{i-1,k}} \} $, (ii) for the $E_{yz}$ component, $N(v_{t_{z_{i,k}}}) =  \{ v_{t_{z_{i,k}}},     v_{u_{i,k}},  v_{u_{i-1,k}} \} $, (iii) for the $H_x$  nodes,  $N(v_{u_{i,k}})  =  \{  v_{u_{i,k}},  v_{t_{x_{i,k}}},  v_{t_{x_{i,k-1}}}, v_{t_{z_{i,k}}}, v_{t_{z_{i,k-1}}} \}$, and  (iv) for the $H_z$  nodes  $N(v_{q_{i,k}})  =  \{  v_{q_{i,k}}, v_{t_{x_{i,k}}}, v_{t_{x_{i-1,k}}}, v_{t_{z_{i,k}}}, v_{t_{z_{i-1,k}}} \}$.  Finally, the corresponding edge weights can be derived in a similar way as before by observing the update equations that arise by discretizing  (\ref{eq:TE_1_PML}) -- (\ref{eq:TE_3_PML}).


 
 
 



\section{Numerical Results}\label{sec:Graph_Results}

In this section, we confirm the validity of our graph-driven numerical solution to Maxwell's equations. To this end, first, we carry out a series of numerical experiments collating the simulation results obtained via GEM and a conventional in-house FDTD solver. In addition, we compare the results of the two solvers for two realistic applications; one from the field of biomedical engineering and one from photonics. Finally, we demonstrate the computational efficiency of GEM to that of the conventional in-house FDTD  implementation with CPU parallelization and another open-source FDTD suite.

\begin{figure*}[t]
\centering
    {\includegraphics[width = 1.5\columnwidth]{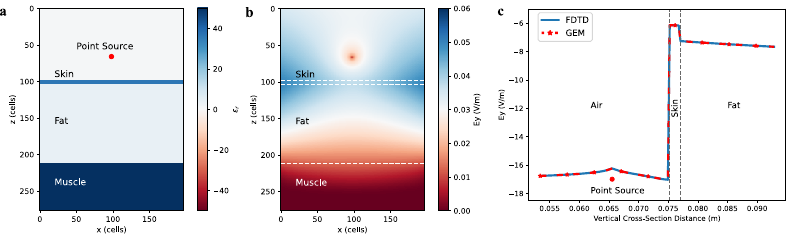}}
\caption{Estimation of SAR with GEM for a biological tissue. \textbf{a}, Electric permittivity for the air, skin, fat, and muscle layers of the biological tissue. \textbf{b}, Snapshot of the electric field intensity dissipated throughout the tissue layers. \textbf{c},   Computed SAR with GEM and FDTD along a vertical cross-section passing through the point source.}
\label{fig:Application_1}
\end{figure*}

\subsection{Proof of GEM Concept}

We verify the validity of our approach by conducting  2,000 simulations with an in-house FDTD package and GEM and juxtaposing the electromagnetic field evolution over space and time for the two methods.  For each simulation, we lay within the simulation grid one or more objects (cylinders, triangles, rectangles, walls) with randomly sampled electromagnetic parameters, and we place a sinusoidal, Gaussian, or modulated Gaussian source at a random position within the grid. We consider rectangular grids with different side sizes of 200, 400, 600, 800, and 1,000 cells, while the number of simulation time steps is selected so that by the end of the simulation the fields have decayed or, in the case of a sinusoidal source, there is a steady-state response. To avoid prolonged simulation times when using the FDTD technique, the number of simulations, $N_{sim}$, is reduced as the grid size increases. Specifically, we conduct 800, 600, 300, 200, and 100 simulations for the grid sizes mentioned previously.

We implement GEM in PyTorch Geometric \cite{Py_Geometric} using an NVIDIA A100 with 80GB of memory. Both GEM and our FDTD implementation use a perfectly matched layer (PML) to avoid artificial reflections at the boundaries of the simulated grid. It is paramount to remark that   GEM does not require training, as its structure is already defined to perfectly mimic the numerical solution to  Maxwell equations. In this sense,  GEM is innately superior to all data-driven approaches proposed for CEM~\cite{karniadakis2021physics,  ML_FDTD_1, ML_FDTD_2, ML_FDTD_3, ML_FDTD_4, ML_FDTD_5, GNNs_FDTD, Dif_Eq_Karniadakis}, since its predictions do not introduce any error, and it employs very simple operations  to infer the electromagnetic fields; specifically, two matrix multiplications and two aggregations. Indicatively, in  Figs.~\ref{fig:Evaluation_Results}a and \ref{fig:Evaluation_Results}b we present several snapshots of the electric field component, $E_y$, computed via GEM and FDTD for the simulated scenarios whose geometry is depicted in Fig.~\ref{fig:Evaluation_Results}c. These qualitative results visually convey how GEM reproduces the propagation of electromagnetic waves in space and time while also capturing their interaction with the environment in a very precise manner. This is further corroborated by Fig.~\ref{fig:Evaluation_Results}d which portrays the waveform recorded at a random point of the simulated domain with the two models, demonstrating once more  perfect correspondence between the results.
From a quantitative standpoint, we compute the coefficient of determination \cite{Coef_Determination}, $R^2$, between the spatiotemporal output of GEM and FDTD across the whole set of experiments conducted, and we find that $R^2=$ 0.9999, clearly indicating the equivalence of the two methods in terms of output quality.


  


\subsection{Applications}

\begin{figure}[t]
\centering
    {\includegraphics[width = 1\columnwidth]{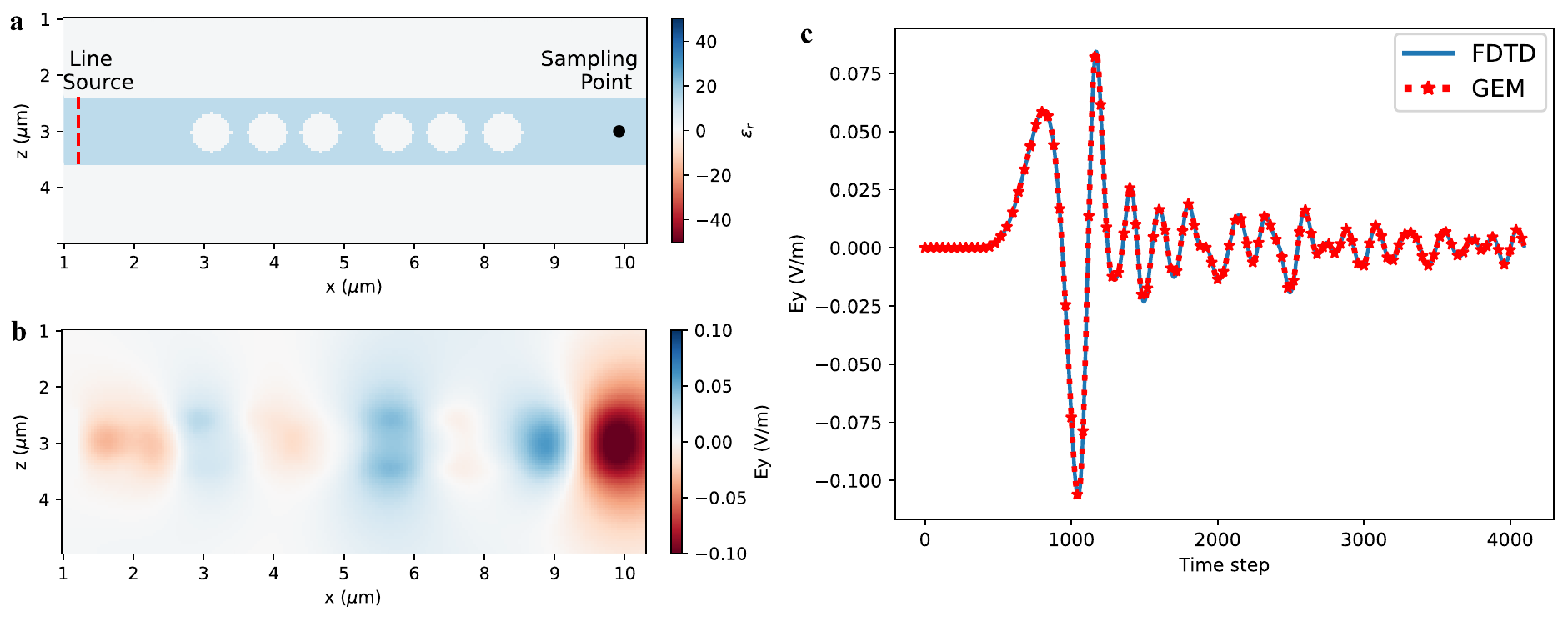}}
\caption{{Wave propagation in a cavity waveguide. \textbf{a} shows the geometry of the waveguide, while \textbf{b} depicts a snapshot of the electric field intensity across the waveguide simulated with GEM. \textbf{c}  Simulated waveform via GEM and FDTD sampled at the end of the waveguide. }}
\label{fig:Application_2}
\end{figure}

To showcase the wide applicability of GEM, we consider two distinct examples, one from the field of computational biology and the other from photonics. The first is to measure the electric field intensity and specific absorption rate (SAR) distribution of microwaves interacting with a biological tissue \cite{App1_Skin_Fat}. We consider a four-layer structure, representing an air-skin-fat-muscle interface. The skin layer has a thickness of 2 mm, that of the fat layer is 33 mm, whilst the muscle layer occupies the remaining space below the fat layer until the end of the grid. The electromagnetic properties of each layer constituting the biological tissue are set according to \cite{App1_Skin_Fat_EM_Parameters}. A Gaussian point source excitation is placed 10 mm above the skin,  with the frequency of the propagation pulse set to 6.8 GHz. The spatial grid is discretized using 20 points per the shortest wavelength of the structure, \ie that occurring in the muscle, and the Courant limit is set to 0.95. A snapshot of the electric field intensity diffused throughout the biological tissue is shown in Fig.~\ref{fig:Application_1}b, while in Fig.~\ref{fig:Application_1}c we present the logarithmic SAR distribution on a vertical cross-section passing through the source estimated via GEM and conventional FDTD. Evidently, the results of the two methods are identical and yield the same SAR, which exhibits a sharp increase when the electromagnetic waves penetrate the skin and subsequent attenuation while propagating through the fat layer.

The second application is emulating wave propagation in a 2D photonic crystal, implemented by simulating a THz Gaussian pulse traversing a periodic dielectric waveguide which assumes a 1D periodicity   \cite{App2_Photonics}. The simulated structure is shown in Fig.~\ref{fig:Application_2}a; the waveguide material has a dielectric constant $\epsilon =12 $, and exhibits a periodicity imposed by two pairs of round cavities with radius $r_c = 0.36 \ {\rm  \mu m}$ spaced  $1 \ {\rm  \mu m}$ apart from each other. The two cavity pairs are separated by  $ 1.4 \ {\rm  \mu m}$ and each one of them comprises 3 circular holes. A line source is placed at the beginning of the simulated grid by exciting every grid cell along the vertical direction of the waveguide with a 75 THz Gaussian pulse. The snapshot of the electric field component shown in Fig.~\ref{fig:Application_2}b illustrates how light is trapped within the waveguide due to the existence of the periodic structure. The sampled waveform shown in Fig.~\ref{fig:Application_2}c evinces again a full equivalency between the graph-driven and the conventional approach.

\begin{figure*}[t]
\centering
    {\includegraphics[width = 1.5\columnwidth]{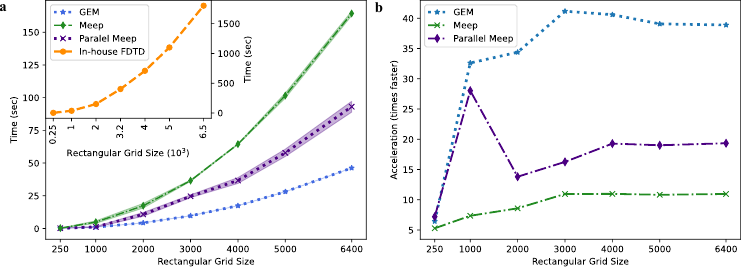}}

\caption{ Performance comparison. \textbf{a}, Simulation time for GEM, our in-house FDTD code and Meep \cite{Meep}. \textbf{b}, Performance gain achieved by GEM and Meep compared to the in-house FDTD code with CPU parallelization.   }
\label{fig:Performance_Comparison}
\vspace*{-8pt}
\end{figure*}

\subsection{Computational Performance Evaluation}

To highlight the expediency of our graph-driven approach we probe the impact of the simulated domain size on the computational time. To this end, in a free space scenario, we gradually increase the size of the simulated domain considering rectangular grids with a side equal to 250, 1,000, 2,000, 3,000, 4,000, 5,000, 6,400 cells, and to quantify the uncertainty in the execution time we run 5 simulations for each cell size, lasting 1,024 time steps.
In addition to comparing the efficiency of GEM to that of our in-house  FDTD code with CPU parallelization, we also benchmark it against a publicly available FDTD suite, namely Meep developed at MIT \cite{Meep}. We note that our in-house FDTD is the standard implementation of the method without leveraging any laborious optimization techniques. On the other hand, Meep constitutes a more sophisticated approach, differing from typical FDTD implementations. Specifically, Meeps enables parallelization by exploiting symmetries, by efficiently storing auxiliary fields only in certain regions (\eg boundary absorbing layers), and by splitting the computational domain into chunks which are combined into an arbitrary topology via boundary conditions, as discussed in \cite{Meep}. Furthermore, Meep also supports distributed-memory parallelism over multi-node clusters,   which we refer to as Parallel Meep, and in the discussion below the simulations are distributed over 10 different processes in a server with  64 CPUs, and 315 GB of RAM.


In Fig.~\ref{fig:Performance_Comparison}a we present the time required to conduct a single FDTD simulation along with the $95 \%$ confidence intervals. As can be observed, the computational time scales exponentially as the grid size increases. Remarkably, for the largest grid size considered, a single CPU-based FDTD simulation with our in-house code takes approximately half an hour, as shown in the inset plot. This time is reduced to approximately 3 and 2 minutes by Meep with and without parallelization, respectively. On the other hand, GEM favored by its native GPU implementation can complete the simulation in less than a minute, \ie almost 40 faster than the standard FDTD implementation and 2 to 4 times faster than an existing highly elaborate state-of-the-art open-source software. Note that currently, GEM   does not make use of any FDTD optimization techniques, \eg grid chunking or exploiting symmetries, employed in Meep.   However, in the future similar functionalities can be embedded in GEM as well, further enhancing its efficiency.  Yet, even without leveraging them, GEM can outperform a state-of-the-art CEM solver in terms of speed. Additional insight regarding the acceleration achieved by the different benchmarks over our in-house FDTD code can be provided by inspecting Fig.~\ref{fig:Performance_Comparison}b.  It can be seen, that as the grid size increases the gain over the conventional FDTD technique plateaus, with   GEM, Meep, and Parallel Meep being approximately 40, 20, and 10 times faster than it, respectively.

 \section{Conclusion}\label{Sec:Disc_and_Conc}

Our work tackles an important problem of electromagnetic modeling: how to train artificial intelligence models to comprehend the laws of electromagnetism and emulate electromagnetic field evolution over space and time. Our simple yet counter-intuitive insight is that there is no need to do so, since there exists a network of neurons, with specific connections and weights, that yields a solution that is equivalent to that of the discretized Maxwell equations. Consequently, the exchange of messages between the network neurons resembles the spatiotemporal evolution of the electromagnetic fields. Unlike data-driven CEM approaches, which are ill-fated owing to the fast error accumulation in their predictions, GEM  yields no error due to the proper setting of the connections between the graph nodes and the corresponding edge weights. The minimal computational complexity entailed by the properly tuned MPNNs along with their native GPU implementation leads to a 40$\times$  acceleration over a direct implementation of the standard FDTD technique with CPU parallelization, while it also outperforms in terms of efficiency previous elaborate state-of-the-art transient response solvers.

The benefits stemming from our work are not confined to the field of physics but they can have a substantial impact on many other sciences in which it is necessary to capture the interaction of electromagnetic fields with matter, such as biomedical engineering and photonics. Furthermore, PDEs describe a plethora of physical phenomena, \eg the Navier–Stokes equations which depict the motion of viscous fluid substances, or the Pennes Bioheat equation which captures the heat transfer between tissue and blood. We are confident that similar GNN structures exist for these problems as well, and we thus advocate pursuing graph-driven approaches where PDE systems are solved via neuron message passing. Ultimately, our heralds a new class of graph-driven numerical solvers,  which outperform existing legacy approaches, and in conjunction with data-driven methods can set the foundation for numerical methods operating outside stability requirements.

\section*{Acknowledgment}

 The work of Stefanos Bakirtzis is supported by the Onassis Foundation and the Foundation for Education and European Culture.

\printbibliography

\end{document}